\documentclass[aps,pre,superscriptaddress,amsmath,amssymb,amsfonts,twocolumn,showpacs]{revtex4-2}
\usepackage[utf8]{inputenc}
\usepackage{amsmath}
\usepackage{amssymb}
\usepackage{amsfonts}
\usepackage{amstext}
\usepackage{amssymb}
\usepackage{amsthm}
\usepackage{mathtools}
\usepackage{physics}
\usepackage{graphicx}
\usepackage{dcolumn}
\usepackage{bm}
\usepackage{braket}
\usepackage{color}
\usepackage[colorlinks=true,citecolor=blue]{hyperref}
\usepackage{natbib}
\usepackage[english]{babel}
\usepackage[table,xcdraw]{xcolor}
\usepackage{setspace}
\usepackage{appendix}
\usepackage{svg}
\DeclareUnicodeCharacter{2212}{\ensuremath{-}}

\definecolor{deeppurple}{rgb}{0.7, 0, 0.8}

\allowdisplaybreaks
\advance\parskip 0.4pt

\begin{document}
\setstretch{1.08}
\title{Spin-mixing-induced dynamics of spinor solitons in F=1 Bose–Einstein condensates}

\author{T. Panagos}
\affiliation{Center for Optical Quantum Technologies, Department of Physics,
University of Hamburg, Luruper Chaussee 149, 22761 Hamburg,	Germany}
\author{A. Romero-Ros}
\affiliation{Departament de Física Quàntica i Astrofísica (FQA), Universitat de Barcelona (UB), c. Martí i Franqués, 1, 08028 Barcelona, Spain}
\affiliation{Institut de Ciències del Cosmos (ICCUB), Universitat de Barcelona (UB), c. Martí i Franqués, 1, 08028 Barcelona, Spain}
\author{G. C. Katsimiga}
\affiliation{Department of Physics and LAMOR, Missouri University of Science and Technology, Rolla, Missouri 65409, USA}
\author{P. Schmelcher}
\affiliation{Center for Optical Quantum Technologies, Department of Physics,
University of Hamburg, Luruper Chaussee 149, 22761 Hamburg,	Germany}
\author{P. G. Kevrekidis}
\affiliation{Department of Mathematics and Statistics, University of Massachusetts Amherst, Amherst, MA 01003-4515, USA}
\affiliation{Department of Physics, University of Massachusetts Amherst, Amherst, 01003-4515, MA, USA}
\affiliation{Department of Mechanical Engineering, Seoul National University, 1 Gwanak-ro, Gwanak-gu, Seoul 08826, South Korea}

\date{\today}

\begin{abstract}
We explore soliton interactions in a homogeneous spinor $F=1$ Bose-Einstein Condensate (BEC) in the presence of a magnetic field, focusing on dark-bright-dark and bright-dark-bright configurations. We investigate how these interactions depend on the phase differences among bright solitons and their influence during the dynamics. Our findings align with prior non-spinor results, i.e., repulsion among in-phase bright solitons and attraction among out-of-phase pairs in self-repulsive atomic BECs.
The potential bright soliton attraction, added to the short-range repulsion of dark-dark soliton interactions, can lead to bound states.
However, we find that these bound states break in the presence of spinor interactions due to the particle exchange dynamics between the hyperfine states of the components. Additonally, we develop an effective classical model to describe the soliton dynamics, using a Lagrangian approach. The accuracy of the model is tested by comparing it against numerical simulations. Our results suggest that the proposed model captures the essential features of soliton behavior in the presence of spin interactions, and provides congruent soliton trajectories and interspecies particle‑exchange dynamics in most of the cases.
\end{abstract}

\maketitle

\section{Introduction} \label{sec:intro}
Among the physical systems that support the rise of nonlinear excitations, Bose–Einstein condensates (BECs) \cite{davisBoseEinsteinCondensationGas1995} occupy a prominent position due to their high controllability, multiple internal states, and long coherence times. In single-component BECs, excitations manifest as matter-wave dark \cite{frantzeskakisDarkSolitonsAtomic2010} and bright \cite{abdullaevDYNAMICSBRIGHTMATTER2012} solitons, corresponding to repulsive and attractive interparticle interactions, respectively. Two-component BECs support more complex structures like dark-bright (DB) solitons \cite{beckerOscillationsInteractionsDark2008, middelkampDynamicsDarkBright2011,hamnerGenerationDarkBrightSoliton2011}, while spinor BECs can host a broader range of dark and bright soliton combinations \cite{nistazakisBrightdarkSolitonComplexes2008,xiongDynamicalCreationComplex2010,bersanoThreeComponentSolitonStates2018,changObservationSpinorDynamics2004,changCoherentSpinorDynamics2005,kawaguchiSpinorBoseEinstein2012,stamper-kurnSpinorBoseGases2013,iedaMatterWaveSolitonsSpinor2004,romero-rosControlledGenerationDarkbright2019}. These multi-soliton structures emerge in repulsively interacting BECs \cite{kevrekidisSolitonsCoupledNonlinear2016}, where bright solitons are supported by the effective potential created by dark solitons—a phenomenon originally noted in nonlinear optics \cite{trilloOpticalSolitaryWaves1988,christodoulidesBlackWhiteVector1988,ostrovskayaNonlinearTheorySolitoninduced1998} and analyzed
in the context of integrable systems~\cite{sheppard}.

Solitons~\footnote{In the present study, we will utilize the terminology of solitons somewhat loosely---as is often
done in physically-minded publications within nonlinear waves: i.e., we will refer to the solitary waves as solitons for brevity
although we recognize them to be coherent structure solutions of a non-integrable field theory.}
exhibit particle-like behavior, allowing for effective descriptions of their motion in harmonic confinements \cite{frantzeskakisDarkSolitonsAtomic2010,kevrekidisSolitonsCoupledNonlinear2016} and interactions between them \cite{katsimigaDarkbrightSolitonInteractions2017,yanMultipleDarkbrightSolitons2011}. Experimental studies have extensively explored soliton dynamics, including collisions of single dark \cite{wellerExperimentalObservationOscillating2008,stellmerCollisionsDarkSolitons2008} and bright solitons \cite{nguyenCollisionsMatterwaveSolitons2014}, as well as the dynamics of two-component \cite{beckerOscillationsInteractionsDark2008}, magnetic \cite{farolfiObservationMagneticSolitons2020}, and three-component solitons \cite{bersanoThreeComponentSolitonStates2018}. The subject continues to be at the forefront of experimental studies, most
recently through the exploration of dense complexes (sometimes referred to also as gases) of dark-bright solitons \cite{mossman2024observation}.

Here, we investigate soliton interactions of a spinor $F=1$ BEC in free space under the influence of a magnetic field, focusing on symmetric dark-bright-dark (DBD) and bright-dark-bright (BDB) configurations. We explore the dependence of the interactions on the phase differences among the bright solitons and examine the influence of spin interactions on the dynamics.
In agreement with previous results  \cite{yanMultipleDarkbrightSolitons2011, katsimigaDarkbrightSolitonInteractions2017}, 
we verify the repulsive (attractive) character of the effective inter-soliton interaction between the bright solitary waves when the latter are in-phase (out-of-phase) in our self-repulsive atomic condensates.
In the attractive case, the formation of bound states is possible.
Yet, spin interactions result in the breaking of the bound state, enabling the solitons to escape. This behavior is associated with significant changes in the individual particle numbers in each component. Furthermore, we propose an effective model based on the Lagrangian approach in order to capture the system's dynamical behavior. Our results indicate that, despite its limitations, the proposed effective model successfully captures the dynamical behavior of the interacting solitons in the majority of cases.

The paper is organized as follows. In Section \ref{sec:model_and_theoretical_setup}, we introduce the physical model — a system of spinor Gross-Pitaevskii equations (SGPEs) — and describe how stationary single DBD and BDB soliton solutions are obtained. Subsequently, we use the single-soliton solutions as building blocks to consider the two-soliton system. Based on the symmetries of the system we identify three different cases, and discuss their properties and how these affect the dynamics. In Section \ref{sec:numerical_results_and_discussion}, we present the numerical results and develop an effective classical approach to describe the dynamics.
Additionally, we provide a comparison between the soliton trajectories and the predictions from the model. The accuracy and limitations of this approach are assessed through fidelity measures comparing model predictions to actual soliton dynamics. Finally, in Section \ref{sec:conclusions}, we summarize our findings and discuss potential future challenges and directions for further research.

\section{Model and theoretical setup} \label{sec:model_and_theoretical_setup}
\subsection{Spinor Gross-Pitaevskii equations}
We consider a 1D spin $F=1$ BEC in free space. In the mean-field approximation, such a system can be described by the following dimensionless spinor Gross-Pitaevskii equations (SGPEs), one for each of the three $m_{F}=0,\pm1$ hyperfine states \cite{kawaguchiSpinorBoseEinstein2012,pitaevskiiBoseEinsteinCondensationSuperfluidity2016}:
\begin{subequations}
    \begin{align}
    i\partial_{t}\Psi_{0} & = -\frac{1}{2}\partial_{x}^2\Psi_{0}
    + g_{n}(|\Psi_{+}|^2 + |\Psi_{0}|^2 +|\Psi_{-}|^2)\Psi_{0} \nonumber \\
    & + g_{s}(|\Psi_{+}|^2 +|\Psi_{-}|^2)\Psi_{0} 
    +2g_{s}\Psi_{+}\Psi_{0}^*\Psi_{-},
    \label{spinor_GPE_0} \\
    \nonumber \\
    i\partial_{t}\Psi_{\pm} & = -\frac{1}{2}\partial_{x}^2\Psi_{\pm} 
    + g_{n}(|\Psi_{+}|^2 + |\Psi_{0}|^2 +|\Psi_{-}|^2)\Psi_{\pm} \nonumber \\
    &+ g_{s}(|\Psi_{\pm}|^2 +|\Psi_{0}|^2 - |\Psi_{\mp}|^2)\Psi_{\pm} 
    +g_{s}\Psi_{\mp}^*\Psi_{0}^2 
    + q\Psi_{\pm}\; ,
    \label{spinor_GPE_pm}
    \end{align}
    \label{spinor_GPE} 
\end{subequations}  
where  $\boldsymbol{\Psi}(x,t) = (\Psi_{+}(x,t) , \Psi_{0}(x,t), \Psi_{-}(x,t))^T$ is the three-component wavefunction, subject to the normalization condition $N =
 \int dx \sum_{m_{F}=-1}^{1} |\Psi_{m_{F}}(x,t)|^2$, where $N$ is the total number of particles.
 In the above equations, $q$ is the quadratic Zeeman parameter, quadratically proportional to an external uniform magnetic field applied along the spin-z direction, and $g_{n}=\frac{a_{0}+2a_{2}}{3}$ and $g_{s}=\frac{a_{2}-a_{0}}{3}$ are the  spin-independent and spin-dependent interaction coefficients, respectively. Here, $a_{0}$ and $a_{2}$ are the s-wave scattering lengths, accounting for two atoms in the scattering channels with total spin $\mathcal{F}=0$ and $\mathcal{F}=2$, respectively. The spin-dependent coefficient is negative for ferromagnetic (F) and positive for antiferromagnetic (AF) spin interactions. 
 Note that Eqs. \eqref{spinor_GPE} were made dimensionless by measuring length, time and energy in units of  $a_{\perp}$, $\omega_{\perp}^{-1}$ and $\hbar\omega_{\perp}$, respectively. Here, $a_{\perp}=\sqrt{\frac{\hbar}{m\omega_{\perp}}}$
is the transverse harmonic oscillator length, corresponding to the experimentally relevant case
\cite{bersanoThreeComponentSolitonStates2018} 
 where a quasi-1D cigar-shaped  system is realized utilizing a highly anisotropic trap with the longitudinal and transverse trapping frequencies obeying $\omega_{x} \ll \omega_{\perp}$. However, it must be emphasized that in our work, we strictly consider a homogeneous 1D system in free space, i.e., $\omega_{x} = 0$. Such possibilities have been explored
 recently in 1D settings in the realm of box potentials~\cite{tamura2025observationmanybodycoherencequasionedimensional}.

Eqs. \eqref{spinor_GPE} conserve the energy, the total number of particles, and the total magnetization along the spin-$z$ axis, $M_{z}=\frac{1}{N}\int dx (|\Psi_{+}|^2 - |\Psi_{-}|^2 )$. However, due to the existence of spin interactions, and more specifically the spin-mixing terms $2g_{s}\Psi_{+}\Psi_{0}^*\Psi_{-}$ and $g_{s}\Psi_{\mp}^*\Psi_{0}^2$, the individual particle numbers of each component, $N_{m_{F}} \equiv
\int dx |\Psi_{m_{F}}(x)|^2$, are not conserved, unless $g_{s}=0$. In the latter case, the spin degrees of freedom are said to be``frozen", and thus throughout this work we will refer to it as non-spinor (NS). Moreover, the absence of such a conservation
will be shown to play a significant role in the full spinorial dynamics.

In case of the spinor systems, any static solution of Eqs. \eqref{spinor_GPE} must fulfill the condition \cite{nistazakisPolarizedStatesDomain2007}: 
\begin{equation}                            
    \Delta \phi \equiv 2\phi_{0} - \phi _{+} - \phi _{-} = 0 \; \text{or} \;  \pi,
    \label{phase_existence_condition}
\end{equation}
where $\phi_{\pm, 0}$ are the phases of the wavefunctions $\Psi_{\pm,0}$. More specifically, $\Delta \phi = 0$ corresponds to the F-system, while $\Delta \phi = \pi$ corresponds to the AF-system. This stems from the fact that a stationary state is the one that minimizes locally the energy functional. With respect to $\Delta \phi$ this is achieved by minimizing the phase-sensitive spin-mixing term 
\begin{align}
    \nonumber
    E_{\text{sm}} &= 2g_{\text{s}} \int dx \Re{\Psi_{\text{0}}^2 \Psi^{*}_{-}\Psi^{*}_{+}} \\ 
    &= 2g_{\text{s}} \int dx [(|\Psi_{-}||\Psi_{+}||\Psi_{\text{0}}|^2)\cos(\Delta \phi). 
    \label{spin_mixing_energy}
\end{align}
It is straightforward to see that $E_{\text{sm}}$ becomes minimal when $\Delta \phi = 0$ ($\Delta \phi = \pi$) in case of a F- (AF-) system.

\subsection{Soliton solutions}
\label{soliton_solutions}
Eqs. \eqref{spinor_GPE} with $g_{s}=0$ (NS-system) reduce to the so-called Manakov model \cite{tsuchidaCoupledModifiedKortewegde1998, iedaExactSolitonSolutions2006}, in particular in its three-component
form which has been explored also experimentally in~\cite{lannigCollisionsThreeComponentVector2020}.
This system of equations is integrable and admits exact vector soliton solutions \cite{yanMultipleDarkbrightSolitons2011, katsimigaDarkbrightSolitonInteractions2017, katsimigaStabilityDynamicsDarkbright2017, katsimigaDarkbrightSolitonPairs2018}.
We consider solutions of the dark 
\begin{equation} 
    \Psi_{d}=\left[\nu\tanh{(D\Tilde{x})} + i\lambda \right]e^{-i\mu_{d} t},
    \label{dark_soliton}
\end{equation}
and bright
\begin{equation} 
    \Psi_{b}=\eta \sech{(D\Tilde{x})}e^{ i\upsilon x -i\mu_{b} t+ i\phi(t)}
    \label{bright_soliton}
\end{equation}
type.
Here $\nu,\eta$ are the amplitudes of the dark and bright components, respectively.
$D$ is a parameter characterizing the common inverse width.
$\Tilde{x}(t)=x-x_{0}(t)$  reflects the spatial dependence,
with $x_0(t)$ representing the instantaneous position of the
soliton center and
$\upsilon= \dot{x}_{0}(t)$ is the soliton velocity.
$\mu_{d}$ and $\mu_{b}$ are the corresponding chemical potentials, and $\phi$ is the bright soliton phase.
$\lambda$ is related to the dark soliton velocity and $\nu^2 + \lambda^2 = n_0$, where $n_0$ is the constant background density of the dark component.

In what follows we will consider vector solitons of the form dark-bright-dark (DBD) and bright-dark-bright (BDB), with the letters denoting the $(+1, 0, -1)$ hyperfine components notation ordering. The aforementioned soliton parameters are connected through 
\begin{align}
    D^2&=\nu_{+}^2+\nu_{-}^2-\eta_{0}^2 \,, \\ 
    \upsilon &=D\frac{\lambda_{+}}{\nu_{+}}= D\frac{\lambda_{-}}{\nu_{-}} \,, \nonumber
    \label{width_amplitude_DBD}
\end{align}
in the DBD case, and
\begin{align}
    D^2 &=\nu_{0}^2-\eta_{+}^2-\eta_{-}^2 \,, \\
    \upsilon &=D\frac{\lambda_{0}}{\nu_{0}} \,, \nonumber
    \label{width_amplitude_BDB}
\end{align}
in the BDB case, while the phase of the bright soliton evolves as
\begin{equation}
    \phi(t) = \frac{1}{2}(D^2 - \upsilon^2)t + (\mu_b - \mu_d)t \,.
    \label{phase_evolution_NS}
\end{equation}
Furthermore, we will consider symmetric configurations, namely  $\nu_{+} = \nu_{-} = \nu$ and $\eta_{+} = \eta_{-} = \eta$, respectively.
Under this assumption, the analytical static solutions of the integrable NS-system for $\upsilon = 0$ read
\begin{equation}
    \boldsymbol{\Psi}=
    \begin{pmatrix}
        \sqrt{\frac{\mu-q}{2}}\tanh{(D\Tilde{x})}   \\
	\sqrt{(\mu-q-D^2)}\sech(D\Tilde{x}) \\ 
	\sqrt{\frac{\mu-q}{2}}\tanh{(D\Tilde{x})}
    \end{pmatrix},
    \label{eq:ansatz_DBD} 
\end{equation}
in the DBD case, and
\begin{equation}
    \boldsymbol{\Psi}=
    \begin{pmatrix}
        \sqrt{\frac{1}{2}(\mu -D^2)}\sech(D\Tilde{x})   \\
        \sqrt{\mu}\tanh{(D\Tilde{x})}  \\ 
        \sqrt{\frac{1}{2}(\mu -D^2)}\sech(D\Tilde{x})
	\end{pmatrix},
	\label{eq:ansatz_BDB} 
\end{equation}
in the BDB case. Note here, that we have set $\mu_{0,\pm}=\mu$. The latter choice has been made with an eye toward the F- and AF-systems, where $N_{m_{F}}$ is not conserved.
Therefore only an overall chemical potential can be defined. Eqs.~\eqref{eq:ansatz_DBD} and \eqref{eq:ansatz_BDB} are used as an ansatz.
Note that in some cases we will include a phase term, $e^{i\phi_{m_F}}$, in order to fulfill \eqref{phase_existence_condition}.
On the numerical side, we used a fixed-point iteration scheme, based on Newton’s method \cite{kelleySolvingNonlinearEquations2003}, to obtain the stationary solutions for the F-, NS- and AF-systems.

\subsection{Two soliton system}
\label{two_soliton_system}
In order to investigate the interactions between two DBD as well as two BDB solitons, we initialize static single solitons of these types, centered at $x_{0}^L=-\frac{d_{0}}{2}$ and $x_{0}^R=\frac{d_{0}}{2}$, where $d_{0}$ is the initial distance between the soliton centers. Then we construct the soliton pair by adding up the bright components and multiplying and normalizing the dark components.

At this point, it should be noted that the aforementioned recipe can be considered a good approximation only for marginally overlapping solitons,
i.e., $d_{0}\gg D^{-1}$. Therefore, in this work, we restrict ourselves to studying only those systems which initially fulfill that condition, and particularly we set the condition $d_{0}>4 D^{-1}$, so as to ensure that the separation of the solitons is much larger than their individual
widths.

In what follows, we investigate the cases where the bright solitons are in-phase or out-of-phase. In order to realize the aforementioned cases, the phases of the	single soliton states are modified 
by acting with operators of the form $\boldsymbol{\hat{T}}_{\theta}=\text{diag}(e^{i\theta_{+}},\;e^{i\theta_{0}},\;e^{i\theta_{-}})$ on them, under the condition that the resulting states fulfill again condition \eqref{phase_existence_condition} and ensure they are stationary solutions.
They are then combined to form the initial two-soliton wavefunctions, which read
\begin{equation}
\boldsymbol{\Psi}(t=0)=
\begin{pmatrix}
\sqrt{\frac{2}{\mu-q}} \Psi_{+} ^{L}\Psi_{+} ^{R} \\
\Psi_{0} ^{L} + \Psi_{0} ^{R} \\ 
\sqrt{\frac{2}{\mu-q}} \Psi_{-} ^{L}\Psi_{-} ^{R}
\end{pmatrix},
\label{initial_state_DBD}
\end{equation}
in case of the two-DBD system, and  
\begin{equation}
\boldsymbol{\Psi}(t=0)=
\begin{pmatrix}
\Psi_{+} ^{L} + \Psi_{+}^{R}  \\
\frac{1}{\sqrt{\mu}} \Psi_{0} ^{L}\Psi_{0} ^{R} \\ 
\Psi_{-}^{L} + \Psi_{-}^{R}
\end{pmatrix},
\label{initial_state_BDB}
\end{equation}
in case of the two-BDB system. In the above states, L and R stand for a left-centered ($x_{0}=-\frac{d_{0}}{2}$) and right-centered ($x_{0}=\frac{d_{0}}{2}$) soliton, respectively. 

Starting from the two-DBD system, two different cases can be identified: one in-phase (IP) configuration, where $\Delta \phi_{0}^{[LR]} = \phi_{0}^{L} - \phi_{0}^{R} = 0$, and one out-of-phase (OP) configuration, where $\Delta \phi_{0}^{[LR]} = \pi$. Here $\phi_{m_F}^{L}$ ($\phi_{m_F}^{R}$) denotes the phase of the left- (right-) centered soliton occupying the $m_F$ component. The two-BDB system consists of two pairs of bright solitons, one for each of the $m_{F}=\pm1$ components. Combining single BDB solutions, two-soliton states with all possible phase difference combinations can be created. However, due to the symmetries of the system, some of these states are physically equivalent. More specifically, they are connected through a $m_{F}=\pm1$ component exchange transformation, a trivial global phase transformation or a transformation of the form $\boldsymbol{\Psi^{\prime}}=\boldsymbol{\hat{T}}\boldsymbol{\Psi}$, where 
$\boldsymbol{\hat{T}}=\text{diag}(e^{i\theta_{+}},\;e^{i\theta_{0}},\;e^{i\theta_{-}})$.
Recall that the condition $2\theta_{0} - (\theta_{-} +\theta_{+}) = 2k\pi$ ensures that two states solve the same SGPE, and thus are physically equivalent. 
After recognizing these symmetries, the system can be reduced to 3 cases:
(IP) $\Delta\phi_{-}^{[LR]}=\Delta\phi_{+}^{[LR]}= 0$,
(OP)  $\Delta\phi_{-}^{[LR]}=\Delta\phi_{+}^{[LR]}= \pi$, and
(IP/OP) $\Delta\phi_{\{-,+\}}^{[LR]}= \{0,\pi\}$, or  $\Delta\phi_{\{-,+\}}^{[LR]}= \{\pi,0\}$,
where $\Delta\phi_{\pm}^{[LR]}=\phi_{\pm}^{L}-\phi_{\pm}^{R}$.

At this point, it is important to note that, since the left and right dark solitons occupying the same component share a common background, the phases of the individually created single solitons are added up  over the entirety of space. In some cases this can lead to the violation of the condition \eqref{phase_existence_condition} from the resulting two-soliton system. In the IP and OP cases, this problem can be solved by multiplying the affected component of the two-soliton system with a proper overall phase. However, in the IP/OP case of the two-BDB system, this is not possible.
More specifically, in case of the phase-sensitive F and AF systems, the violation of the above conditions at the right side ($x>0$) and at the left side ($x<0$) of the system, respectively, is unavoidable. This violation excites the affected part of the system.
Therefore, one cannot directly speak of an interaction between two solitons; instead, it pertains to the interaction between a soliton and an excited state. 
Hence, the IP/OP case will not be included in the present study, since it does not correspond to a true soliton-soliton interaction. 

Our goal is to study how  the interaction depends on the phase difference between the bright solitons.
Thus, it is reasonable to anticipate that the initially created phase differences will be conserved during the dynamics. We will show that, in both IP and OP cases, this assumption is fulfilled. 
First, it is important to note that in these cases, the initial states are symmetric with respect to the $m_{F} = \pm 1$ components; specifically, $\Psi_{+}(t=0) = \Psi_{-}(t=0) $. Given that Eqs. \eqref{spinor_GPE} are symmetric under the exchange of the $m_{F} = \pm 1$ components, this symmetry will be conserved over time, ensuring that $\Psi_{+}(t) = \Psi_{-}(t)$. Therefore Eqs. \eqref{spinor_GPE} reduce to
\begin{subequations}
    \begin{align}
    i\partial_{t}\Psi_{0} & =-\frac{1}{2}\partial_{x}^2\Psi_{0}
    + g_{n}(|\Psi_{1}|^2 + |\Psi_{0}|^2)\Psi_{0} \nonumber  \\
    &+ g_{s}|\Psi_{1}|^2\Psi_{0} 
    +g_{s}\Psi_{1}^2\Psi_{0}^* ,
    \label{symmetric_spinor_GPE_0} \\
    \nonumber \\
    i\partial_{t}\Psi_{1} & =-\frac{1}{2}\partial_{x}^2\Psi_{1} 
    + g_{n}(|\Psi_{1}|^2 + |\Psi_{0}|^2)\Psi_{1} \nonumber \\
    &+ g_{s}|\Psi_{0}|^2\Psi_{1} 
    +g_{s}\Psi_{1}^*\Psi_{0}^2 
    +q\Psi_{1},
    \label{symmetric_spinor_GPE_pm}
    \end{align}
    \label{symmetric_spinor_GPE} 
\end{subequations}
where $\Psi_{1}=\sqrt{2}\Psi_{\pm}$. A key feature of the above equations is that they conserve the parities of each of the $\Psi_{0}$ and $\Psi_{1}$ components individually, if the latter have initially definite (but not necessarily equal) parities. 
The initial two-soliton states of the cases fulfill by construction this condition. More specifically, the dark components have even parity, while the bright ones in the IP (OP) configuration have even (odd) parity. The connection between the phase differences $\Delta \phi_{0}^{[LR]} = 0,\;\pi$ and the parity of the corresponding wavefunctions becomes clear if one considers that the phase of every complex function $f(x)=|f(x)|e^{i\phi(x)}$ with even parity fulfills $\phi(x)-\phi(-x)=2k\pi$ ($k=0,1,2,...$), while in case of a function with odd parity the phase fulfills $\phi(x)-\phi(-x)=(2k+1)\pi$ ($k=0,1,2,...$). Consequently, the parity-conservation leads to phase difference conservation.

\section{Numerical results and discussion} \label{sec:numerical_results_and_discussion}
In this section we present the dynamics of the aforementioned states comparing the F-, NS-, and AF-systems. For this purpose we set the chemical potential to $\mu = 2$ (without loss of generality), the spin-independent interaction coefficient $g_{n}=1$, and the spin-dependent one $g_{s}=-5\times 10^{-3}$, $g_{s}=0$, and $g_{s}=5\times 10^{-3}$ in case of F- NS- and AF-system, respectively. The latter are representative values in the neighborhood of realistic ones for $^{87}$Rb and $^{23}$Na \cite{PhysRevA.75.023617}. Since our goal is to investigate the interactions between different-sized solitons, we vary the quadratic Zeeman term ($q$) within the intervals 
$[-1,-0.1]$ and $[0.1,1]$ in case of a DBD and BDB system, respectively. The motivation to use these values stems from Ref. \cite{katsimigaPhaseDiagramStability2021}, where, among others, the $q$-intervals of stability of trapped single BDB and DBD solitons have been investigated. Our chosen $q$-values lie within the stability intervals. Moreover, we additionally confirmed that they lead to stable soliton solutions in free space as well, which is a key ingredient for our considerations.

For the time evolution of the initial states, a fourth-order Runge–Kutta integrator is utilized, while a second-order finite differences method is used for the spatial derivatives. The spatial and time discretization are $dx=5\times10^{-2}$ and $dt=10^{-3}$, respectively. Convergence of the results has been verified for smaller values of $dx$ and $dt$ for representative selected cases within the initial parameter space. 

\subsection{Dynamics}
\label{IP_and_OP}
\begin{figure}[t]
    \centering
    \includegraphics[width=\linewidth]       {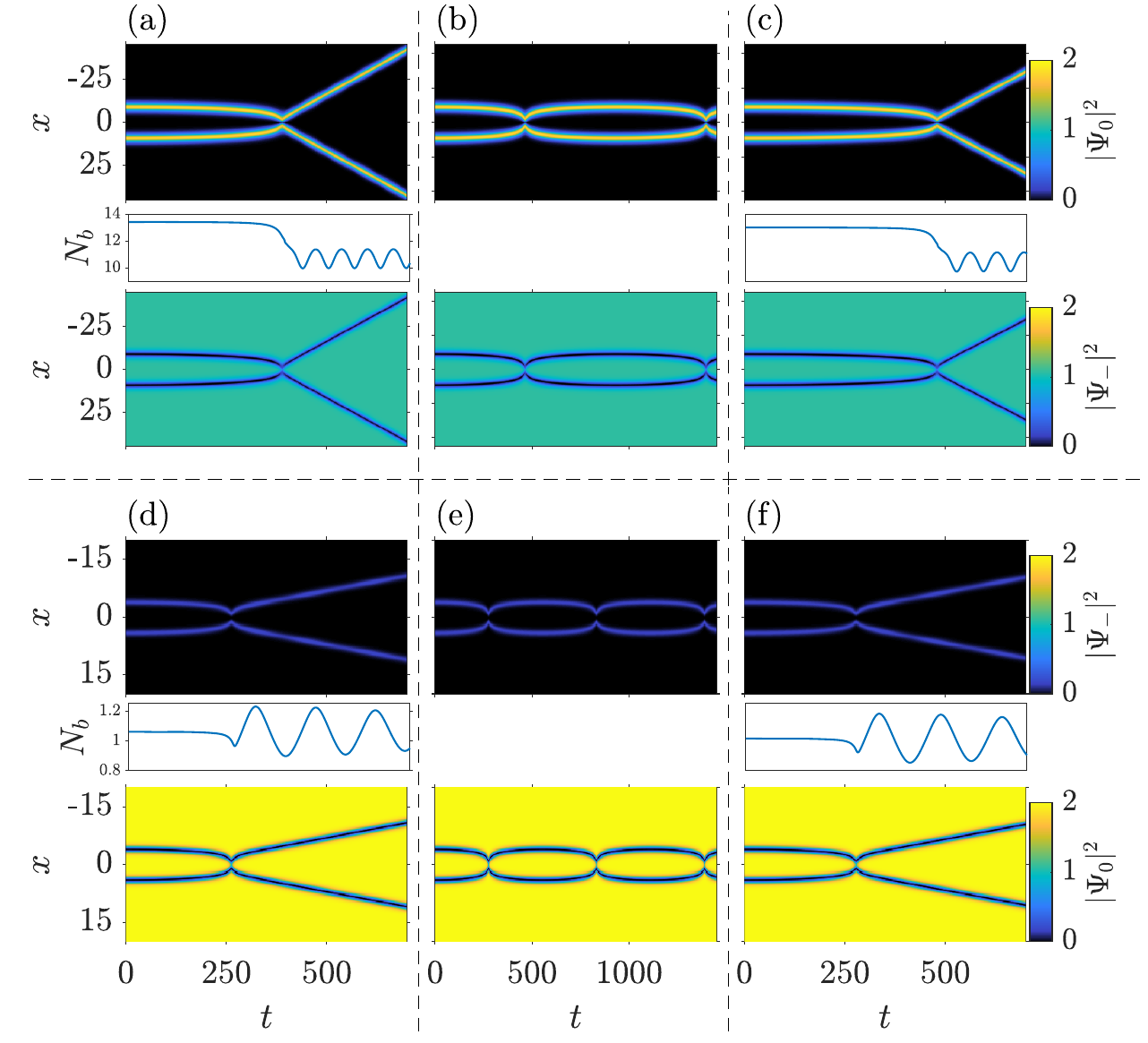}
    \caption{Example of the spatio-temporal evolution of the densities of the dark ($|\Psi_{0}|^2$) and bright ($|\Psi_{-}|^2$) components of an OP two-DBD (two-BDB) [(a)--(c)] ([(d)--(f)]) system in case of the F- [(a), (d)], NS- [(b), (e)] and AF- [(c), (f)] interactions, for $d_{0}=18$ and $q=-0.16$ ($d_{0}=8$ and $q=0.835$) [(a)--(c)] ([(d)--(f)]). In case of the F- and AF- systems, the time evolution of the number of particles of the bright components ($N_{b}$) is depicted as well. The ($|\Psi_{+}|^2$) component is not displayed here, since $\Psi_{+}(t) = \Psi_{-}(t)$, as discussed in Section \ref{two_soliton_system}.
    }
    \label{OP_densities_and_Nb_vs_t}
\end{figure}

In order to investigate the dynamical behavior of the two-DBD as well as the two-BDB system in the IP and OP configurations, we consider the time evolution of the initial two-soliton state \eqref{initial_state_DBD}, in case of the two-DBD system, and \eqref{initial_state_BDB}, in case of the two-BDB system, for varying initial distances and $q$-values. 

For small $|q|$-values both systems exhibit very similar dynamics, which is expected given that Eqs. \eqref{symmetric_spinor_GPE} tend to be symmetric upon exchange of the $\Psi_{0}$ and $\Psi_{1}$ components as $q \rightarrow 0$. For increasing $|q|$-values the differences become visible, although also in this regime the qualitative characteristics of the dynamics remain the same. More specifically, a repulsive (attractive) effective interaction between the left and the right bright solitons occurs when the systems are in the IP (OP) configuration (see Fig. \ref{OP_densities_and_Nb_vs_t}(b) and (e) for the OP configuration of the NS system) in agreement with previous results  \cite{yanMultipleDarkbrightSolitons2011, katsimigaDarkbrightSolitonInteractions2017}. However, in case of the OP configuration, the spinor systems (F and AF) exhibit significantly different behavior from the NS one, that is, instead of forming a bound state they appear to escape after one (or a few) collision(s) (see Fig. \ref{OP_densities_and_Nb_vs_t}(a), (c), (d) and (f)). Note that instances where solitons undergo multiple collisions before escaping are rare and typically occur at shorter initial distances, where the two-soliton ansatz has limited validity.

The dynamics of the OP spinor systems includes also significant changes of the individual particle numbers of the $m_{F}=0,\pm 1$ components. More specifically, as the left and right solitons deform while approaching the collision point, inter-component particle exchange can be identified (Fig. \ref{OP_densities_and_Nb_vs_t}(a), (c), (d) and (f)), associated with the activation of the spin-mixing terms of the Hamiltonian, followed by post - collision  oscillatory behavior, a characteristic signature of the breathing motion of the solitons. Notably, in systems with small $|q|$-values, a permanent inter-component mass transfer occurs, since the post-collision average particle number of the bright component $N_{b}$ lies clearly below the initial value (Fig. \ref{OP_densities_and_Nb_vs_t}(a) and (c)). The described spin-mixing-induced particle exchange is a central feature of the spinorial system whose role, to the best of our knowledge, has not been previously appreciated.

\subsection{Effective description}
In what follows we attempt to capture the dynamical behavior of the above studied systems, and more importantly the ability of the OP spinor solitons to escape, adopting an effective classical particle approach. More specifically, we employ the adiabatic approximation of
the Lagrangian theory for solitons, according to which, the soliton parameters become slowly varying functions of time, but the functional form of the soliton remains unchanged \cite{kivsharLagrangianApproachDark1995}. 

In order to build the two-soliton ansatz, we recall Sec. \ref{two_soliton_system} where we have shown that, in case of the IP and OP systems, Eqs. \eqref{spinor_GPE} can be reduced to Eqs. \eqref{symmetric_spinor_GPE}. Therefore, a DBD (BDB) system reduces effectively to a DB system with $\mu_{d}=\mu_{1}=\mu - q$, $\mu_{b}=\mu_{0}=\mu$ ($\mu_{d}=\mu_{0}=\mu$, $\mu_{b}=\mu_{1}=\mu - q$)
Note that the quadratic Zeeman term $q$ acts as an effective chemical potential on the $\Psi_{1}$ component. After rescaling space-time coordinates as $t \rightarrow \mu_d t$, $x \rightarrow \sqrt{\mu_d}x$ and the densities $|\Psi|^2 \rightarrow \mu_d^{-1}|\Psi|^2$, the ansatz for the dark and bright components of the two-soliton state reads 
\begin{align}
    \nonumber
    \Psi_d(x,t) &= \left(\nu \tanh(D\Tilde{x}_{-}) + i \lambda \right) \left( \nu \tanh(D\Tilde{x}_{+}) - i \lambda\right) \\
    \nonumber
     \Psi_b(x,t) &=\frac{ \sqrt{DN_b}}{2}\sech(D\Tilde{x}_{-})e^{i\left(x\frac{\dot{\xi}}{2} + \varphi(t)\right)} \\ 
     & \, \,\,\,\,\,\, + \frac{ \sqrt{DN_b}}{2}\sech(D\Tilde{x}_{+})e^{i\Delta \theta}e^{i\left(- x\frac{\dot{\xi}}{2} + \varphi(t)\right)},
\end{align}
where $\Tilde{x}_{\pm}(t) = x \pm \frac{\xi(t)}{2}$, with $\xi(t)$ being the distance between the soliton centers, and $\Delta \theta$ the phase difference between left and right bright solitons of the same hyperfine component.
Within the adiabatic Lagrangian approach, the soliton parameters become time dependent variational parameters. The Lagrangian density reads 
\begin{align}
\nonumber
    \mathcal{L}\{ \Psi \} = \, &\frac{i}{2}(\Psi^{*}_d \partial_t \Psi_d - \Psi_d \partial_t \Psi^{*}_d)(1 - \frac{1}{|\Psi_d|^2}) \\
    \nonumber
    &+ \frac{i}{2}(\Psi^{*}_b \partial_t \Psi_b - \Psi_b \partial_t \Psi^{*}_b) \\ 
    \nonumber
    &- \frac{1}{2}|\partial_x \Psi_d|^2 - \frac{1}{2}|\partial_x \Psi_b|^2 \\
    \nonumber
    &- \frac{1}{2}(|\Psi_d|^2 + |\Psi_b|^2 - 1)^2 + (\Tilde{\mu} - 1)|\Psi_b|^2 \\
    &- g_s \left(|\Psi_d|^2|\Psi_b|^2 + Re\{(\Psi^{*}_d)^2\Psi^2_b\}\right),
     \label{Lagrangian_density}
\end{align}
from which the averaged Lagrangian is obtained
\begin{equation}
L= \int_{- \infty}^{\infty}\mathcal{L} \, dx=\mathcal{T} - (2 E_1 + E_{\text{dd}} + E_{\text{bb}} + E_{\text{db}} + E_{\text{s}}),
\label{averaged_Lagrangian}
\end{equation}
where

\begin{subequations}
\begin{align*}
\mathcal{T} \approx \,&
    2 \dot{\xi}\left(- \nu \lambda + \arctan(\frac{\nu}{\lambda}) \right)
    - \frac{N_b}{4}\xi \, \ddot{\xi} \\
& - N_b \dot{\varphi} \left(1 + 2 D \cos(\Delta\theta) \xi 
 \,e^{-D\xi} \right) \\
& - \frac{4 \nu \dot{\lambda}}{D} \left(1 + 2(1 - 2 D\xi)
 \,e^{-2D\xi}\right),
\\[1.0em]
E_1 = & \,\frac{2\nu^2}{3}(D - \frac{N_b}{2} + \frac{\nu^2}{D})
 + \frac{N_b}{16}\dot{\xi}^2  \\
& +\frac{D N_b}{12}(D + \frac{N_b}{2})
 + \frac{N_b}{2}(1-\Tilde{\mu}),
\\[1.0em]
E_{\text{dd}} \approx \,&
 16\left(
  -\frac{7}{3 D}
  + \frac{4 D}{3}
  + (1 - D^2 )\, \xi
\right)
\, e^{-2 D \xi},
\\[1.0em]
E_{\text{bb}} \approx \,&
 D \,N_b
  \left(
    2 D (1 - \frac{1}{2} D \xi)
    +  N_b
  \right)  \cos(\Delta\theta) e^{-D \xi}
\\[1em]
& 
 - D \, N_b^2 \,(1 + 2 \cos^2(\Delta\theta))\, (1 - D \xi)\, 
 e^{-2 D \xi},
\\[1.0em]
E_{\text{db}} \approx \,&
     4 N_b \left(\frac{1}{2}(1 - \Tilde{\mu}) D \xi - 1 \right)
     \cos(\Delta\theta) e^{-D\xi} \\
& + \left( \frac{40}{3} - 8 D \xi \right) N_b\, e^{-2D\xi},
\\[1.0em]
E_{\text{s}} \approx \,&
  \frac{g_s}{3}  N_b\big[
  1
  + \cos(\Delta\theta)
       \cos(\Delta\theta + 2\varphi)
   \big],
\end{align*}
\end{subequations}
and $\Tilde{\mu} \equiv \frac{\mu_b}{\mu_d}$ is the rescaled chemical potential. In the above expression, $\mathcal{T}$ collects all contributions involving time derivatives of the fields, $E_1$ is the (potential) energy of a single DB soliton, $E_{dd}, \,  E_{bb}$ and $E_{db}$ account for the interaction between the two solitons, with $E_{dd}$ denoting the ---generically repulsive--- 
interaction between the two darks, $E_{bb}$ the interaction between the two brights and $E_{db}$ the interaction energy between the dark soliton of one component and the bright soliton in the other component, while $E_s$ accounts for the spin interactions. 
For the bright soliton interaction, we notice the dependence on $\cos(\Delta \theta)$, which crucially changes the nature
of the force as $\Delta \theta$ varies from $0$ to $\pi$.

Since we consider slowly moving, well-separated solitons, and a small spin-coupling parameter $g_s$, we retain only terms up to order $\mathcal{O}(g_s)$, $\mathcal{O}(e^{-2D\xi})$, and $\mathcal{O}(\dot{\xi}^2)$, and neglect higher powers as well as mixed products of these small quantities. Within this approximation, the spin interaction acts as an internal contribution for each soliton, independent of both the soliton–soliton interaction (terms $\mathcal{O}(g_s\,e^{-D\xi})$ and higher are neglected) and the soliton's motion (terms $\mathcal{O}(g_s\,\dot{\xi})$ and higher are neglected). Likewise, the soliton-soliton interaction is independent of the soliton's motion (terms $\mathcal{O}(\dot{\xi} \, e^{-D\xi})$ and higher are neglected).

The variational parameters are the soliton distance $\xi(t)$, and the phase $\varphi(t)$ and particle number $N_b(t)$ of the bright component, while $\lambda(t)$ and $D(t)$ are constrained by the solitonic relations.
\begin{align}
\nonumber
    D^2 = 1 - \lambda^2 - \frac{DN_b}{4}, \\
    \nonumber
    \lambda = \frac{\nu \dot{\xi}}{2D}, \\
    \nu^2 + \lambda^2 = 1.
    \label{solitonic_relations}
\end{align}
For slow moving solitons, the above system of equations can be approximately expressed in terms of the free variational parameters as follows
\begin{align}
\nonumber
    D = -\frac{N_{b}}{8}+\sqrt{1 + \lambda^2 + (\frac{N_{b}}{8})^{2}} \\
     \lambda \approx \frac{\dot{\xi}}{2D^{(0)}},
    \label{approx_of_lambda_and_D}
\end{align}
where 
\begin{equation}
  D^{(0)}=-\frac{N_{b}}{8}+\sqrt{1 + (\frac{N_{b}}{8})^{2}}.
  \label{inverse_width_without_lambda}
\end{equation}
In order to get these approximations, we neglect terms $\mathcal{O}(\dot{\xi}^3)$ and higher. Note here, that, in the absence of soliton-soliton interaction, the phase $\varphi(t)$ and particle number $N_b(t)$ of the bright component form a canonical pair.

The time evolution of the soliton parameters can be obtained by applying the Euler-Lagrange equations
\begin{equation}
    \partial_{\varphi} L - 
\frac{d}{dt}(\partial_{\dot{\varphi}} L)  = 0,
    \label{EL_wrt_phi}
\end{equation}
\begin{equation}
    \partial_{N_b} L - 
\frac{d}{dt}(\partial_{\dot{N_b}} L)  = 0,
    \label{EL_wrt_Nb}
\end{equation}
and
\begin{equation}
    \partial_{\xi} L - 
\frac{d}{dt}(\partial_{\dot{\xi}} L) + 
\frac{d^2}{dt^2}(\partial_{\ddot{\xi}} L) = 0
    \label{EL_wrt_ksi}
\end{equation}
which lead to the evolution equations
\begin{equation}
\dot{N}_b \approx - \frac{2}{3} \, g_s N_b \cos(\Delta\theta) \sin(\Delta\theta + 2\varphi),
    \label{EOM_for_Nb}
\end{equation}
\begin{equation}
    \dot{\varphi} \approx \frac{1}{2}(D^2 - \frac{\dot{\xi}^2}{4}) + \Tilde{\mu} - 1,
    \label{EOM_for_phi}
\end{equation}
and
\begin{widetext}
\begin{align}
\nonumber
 \ddot{\xi} \approx &\,  \bigg[ \frac{8 N_b (D^{(0)})^4 \cos(\Delta\theta)}{8 - D^{(0)} N_b} \bigg]e^{-D^{(0)}\xi} \\ 
&+ \bigg[ \frac{4 D^{(0)} \left(-272 + D^{(0)}\left(96 \xi + 104\,N_b + D^{(0)}\left(176 - 96 D^{(0)}\xi
 - 48 \,N_b \xi + 3\left(2 + \cos(2\Delta\theta)\right)\left(-3 + 2 D^{(0)} \xi\right) N_b^{2}\right)\right)\right)}
     {3\left(-8 + D^{(0)}\,N_b\right)} \bigg]e^{-2D^{(0)}\xi}.
     \label{EOM_for_ksi}
\end{align}
\end{widetext}

Note that terms $\mathcal{O}(g_s)$ and $\mathcal{O}(e^{-D\xi})$ have been neglected from Eq. \eqref{EOM_for_phi}, since the latter is already $\mathcal{O}(1)$ and therefore interaction and spinor terms serve as small corrections. On the other hand, terms $\mathcal{O}(g_s)$ are neglected from Eq. \eqref{EOM_for_ksi} since they would lead to spurious self‑acceleration terms, indicating forces acting on the solitons even at large distances, where the interaction between them is practically negligible. 
At these distances, the solitary waves evolve as individual, independent entities and hence such terms are artifacts of the 
ansatz.

To assess the validity of the suggested model, the system of equations Eq. \eqref{EOM_for_Nb},\eqref{EOM_for_phi} and \eqref{EOM_for_ksi} has been evaluated numerically using a fourth-order Runge–Kutta integrator. The initial parameters are $\xi(t=0) = d_{0}$, where $d_{0}$ is the initial distance between the solitons, $\dot{\xi}(t=0) = 0$, $\varphi(t = 0) = 0$ ($\varphi(t = 0) = \frac{\pi}{2}$) in case of the F- (AF-) system, and $N_b(t=0) = \frac{4\left(1-2(1 - \Tilde{\mu}) \right)}{\sqrt{2(1 - \Tilde{\mu})}}$. The latter is obtained from Eq. \eqref{EOM_for_phi} by imposing $\dot{\varphi} = 0$ in case of a stationary state in order to fulfill Eq.~\eqref{phase_existence_condition} (note that the latter requires fixing $\varphi(t)$ to a specific value). This approach provides  $N_b(t=0)$ values which deviate from the ones taken from the our numerical simulations, in most cases, less than $5\%$.

To quantitatively compare the predicted trajectories $\xi(t)$ with the original soliton trajectories $d(t)$, we introduce the following fidelity measure, based on the sMAPE metric \cite{shuklaSaltRockCreep2025}
\begin{equation} 
    F = 1 - \frac{1}{\mathcal{N}}\sum_{i=1}^{\mathcal{N}}\frac{|\dot{d}(t_{i}) - \dot{\xi}(t_{i})|}{|\dot{d}(t_{i})| + |\dot{\xi}(t_{i})|}.
    \label{Fidelity}
\end{equation}
Note that we use the corresponding velocities instead of the trajectories themselves to avoid the potential contribution of a constant offset and to capture the dynamical character of the matching of the respective trajectories.

For the NS systems in the OP configuration, the number of time points  ($\mathcal{N}$) considered in the evaluation is determined by the criterion that the solitons have experienced at most three collisions. In all other scenarios, $\mathcal{N}$ corresponds to the time point when the soliton acceleration drops below $10^{-4}$, signifying negligible interaction and indicating that the solitons move at constant velocity.

\begin{figure}[t]
    \centering
    \includegraphics[width=\linewidth]{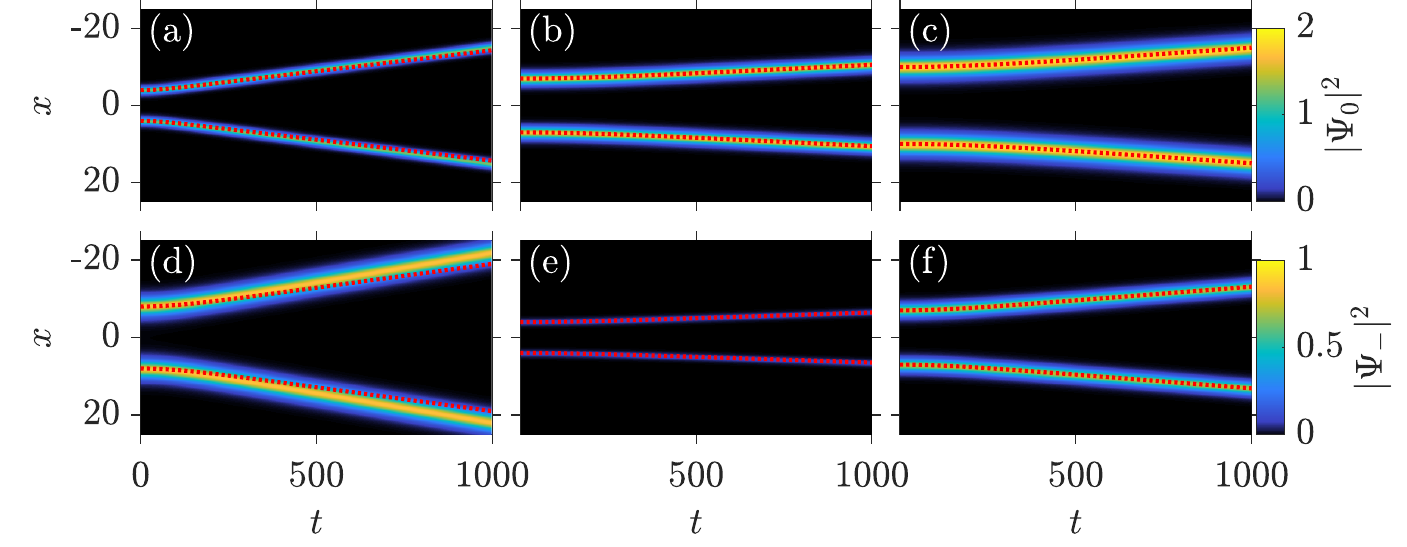}
    \caption{Characteristic results of the comparison between the spatio-temporal evolution of the IP systems and the trajectories predicted from the model (red dotted lines). The depicted cases: 
    (a) two-DBD F-system with $q = -0.72$ and $d_{0}=8$ ($F = 0.96$), 
    (b) two-DBD NS-system with $q=-0.32$ and $d_{0}=14$ ($F = 1$), 
    (c) two-DBD AF-system with $q=-0.12$ and $d_{0}=20$ ($F = 0.98$), 
    (d) two-BDB F-system with $q=0.145$ and $d_{0}=16$ ($F = 0.86$), 
    (e) two-BDB NS-system with $q=0.895$ and $d_{0}=8$ ($F = 1$), 
    (f) two-BDB AF-system with $q=0.26$ and $d_{0}=14$ ($F = 0.99$).
    }
    \label{densities_vs_model_IP}
\end{figure}
 
\begin{figure}[t]
    \centering
    \includegraphics[width=\linewidth]{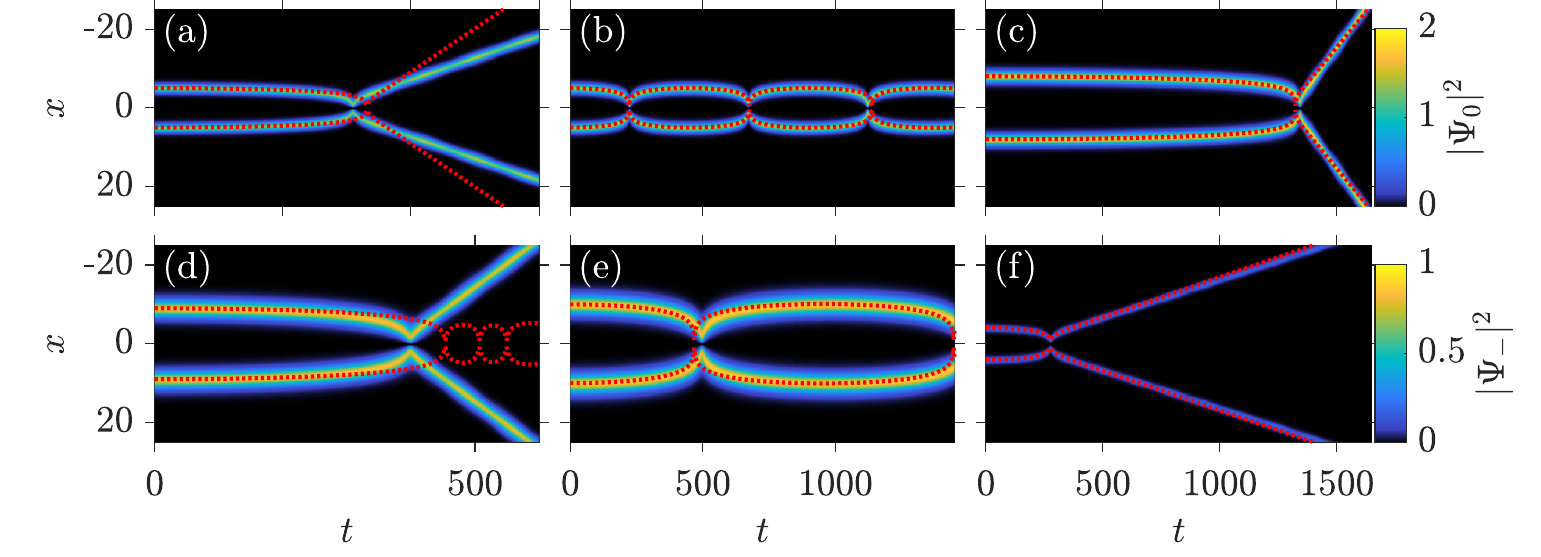}
    \caption{Same as Fig. \ref{densities_vs_model_IP} but for the OP configuration. The depicted cases: 
    (a) two-DBD F-system with $q=-0.67$ and $d_{0}=10$ ($F = 0.71$), 
    (b) two-DBD NS-system with $q=-0.56$ and $d_{0}=10$ ($F = 0.96$), 
    (c) two-DBD AF-system with $q=-0.36$ and $d_{0}=16$ ($F = 0.88$), 
    (d) two-BDB F-system with $q=0.16$ and $d_{0}=18$ ($F = 0$), 
    (e) two-BDB NS-system with $q=0.12$ and $d_{0}=20$ ($F = 0.80$), 
    (f) two-BDB AF-system with $q=0.835$ and $d_{0}= 8$ ($F=0.98$).}
    \label{densities_vs_model_OP}
\end{figure}

\begin{figure}[t]
    \centering
    \includegraphics[width=\linewidth]{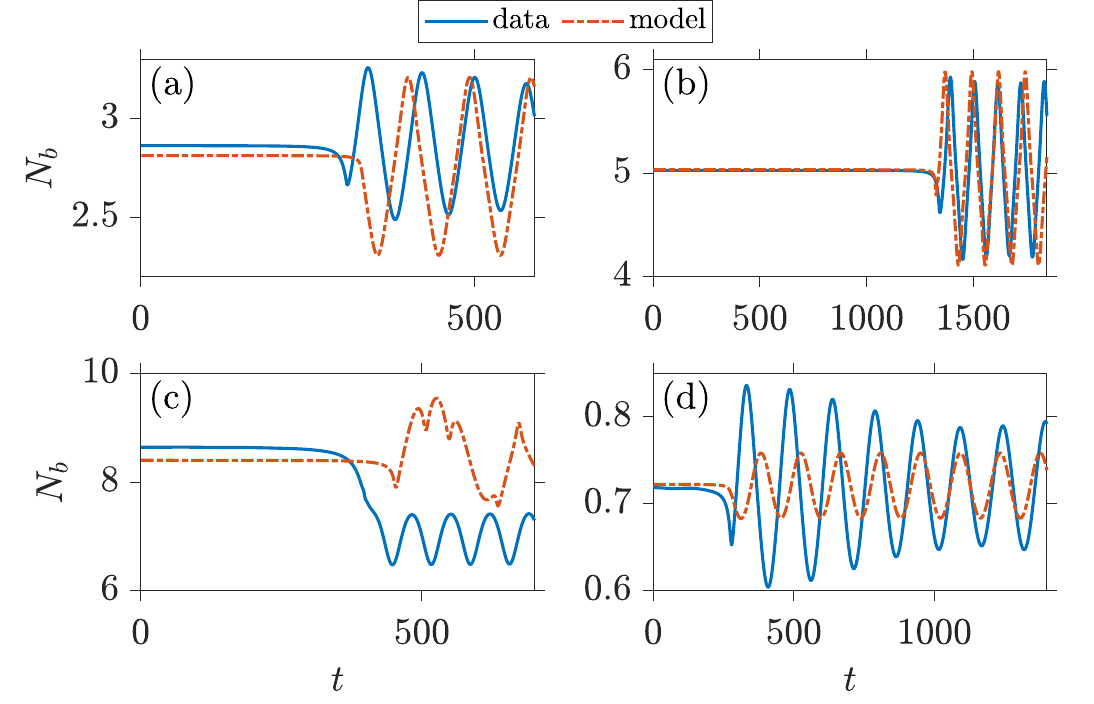}
    \caption{Characteristic results of the comparison between the time evolution of the number of particles of the bright components ($N_{b}$) and the model predictions in case of the OP system. The depicted cases: 
    (a) two-DBD F-system with $q=-0.67$ and $d_{0}=10$, 
    (b) two-DBD AF-system with $q=-0.36$ and $d_{0}=16$, 
    (c) two-BDB F-system with $q=0.16$ and $d_{0}=18$ ($F = 0$), 
    (d) two-BDB AF-system with $q=0.835$ and $d_{0}= 8$.}
    \label{Nb_vs_model_OP}
\end{figure}
 
\begin{figure}[t]
    \centering
    \includegraphics[width=\linewidth]{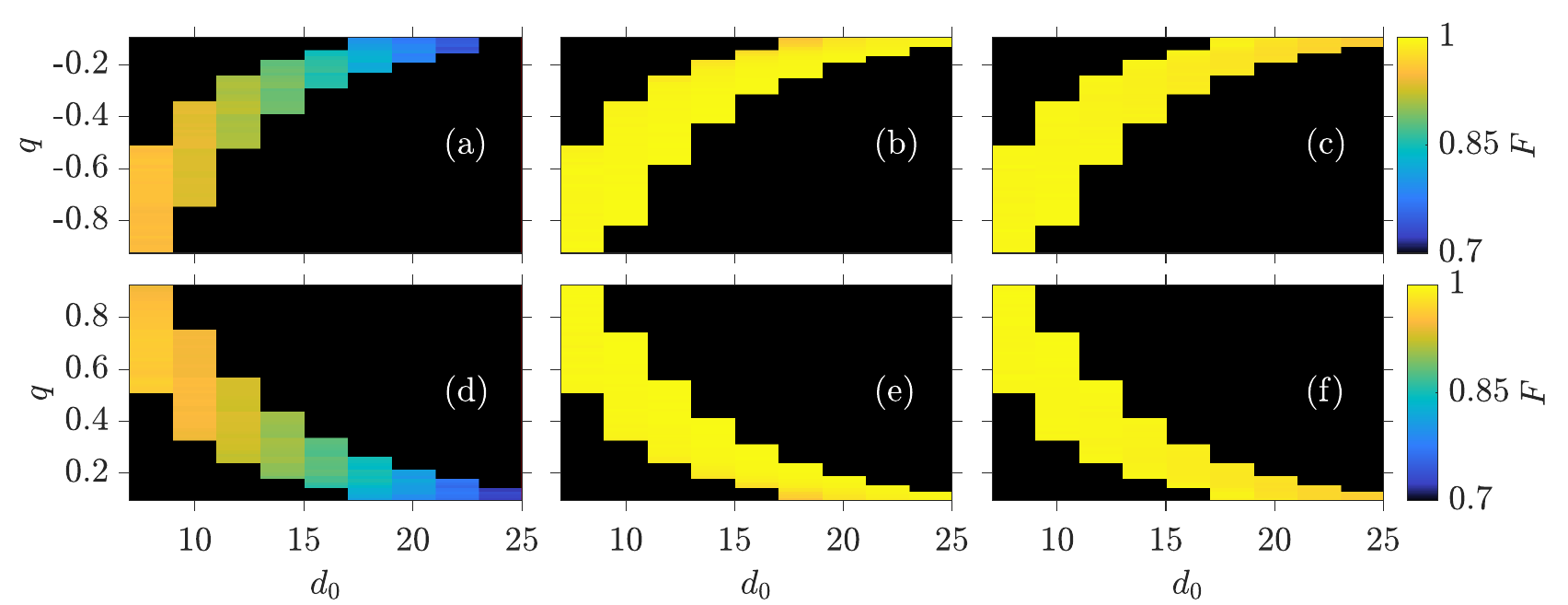}
        \caption{Fidelities of the IP systems: 
        (a) two-DBD F-system, 
        (b) two-DBD NS-system, 
        (c) two-DBD AF-system, 
        (d) two-BDB F-system, 
        (e) two-BDB NS-system, 
        (f) two-BDB AF-system.}
    \label{F_tot_IP}
\end{figure}

\begin{figure}[t]
    \centering
    \includegraphics[width=\linewidth]{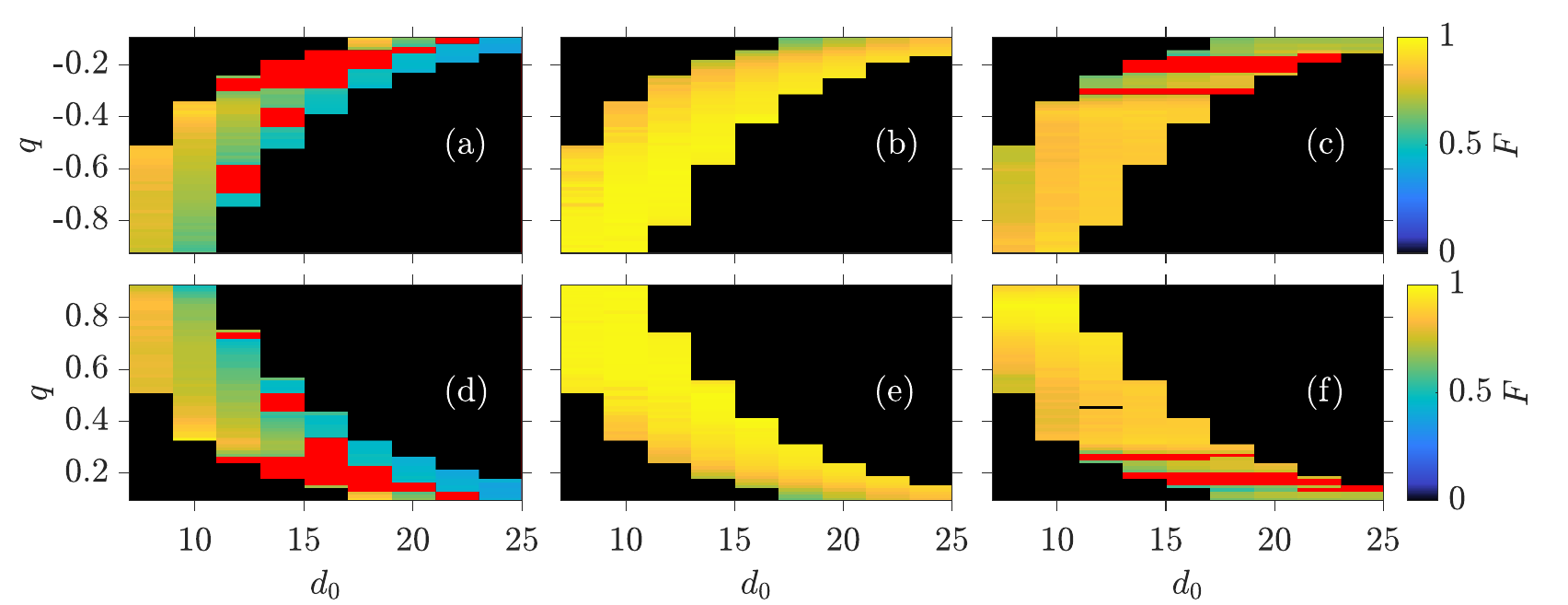}
    \caption{Same as Fig. \ref{F_tot_IP} but for the OP configuration.}
    \label{F_tot_OP}
\end{figure}

In Fig.~\ref{densities_vs_model_IP} (Fig. \ref{densities_vs_model_OP}) we compare the effective model with our numerical simulations for the IP (OP) system. 
In the IP case, we can observe a very reasonable agreement of the solitonic repulsive trajectories within the spinor system. While
the effect of the pairwise solitonic repulsion 
seems to be weakest in the NS case, it appears to be captured  well in all 3 scenarios (F, NS and AF, respectively
in the figure).
In Fig. \ref{Nb_vs_model_OP}, we depict the time evolution of the number of particles of the bright components ($N_{b}$) in case of the spinor OP 
systems. It is once again clear that the exchange of particles between the 
components and the activation of the spin-mixing terms is 
critical for observing the resulting phenomenology.
The aggregated results for the fidelities for both systems and all available initial parameter sets ($d_{0},q$) are shown in Fig.~\ref{F_tot_IP} and Fig.~\ref{F_tot_OP}. 
Note here that we consider the comparison to be unsuccessful in cases involving spinor OP systems where the effective model incorrectly forecasts a bound state within the simulation timeframe. These cases are highlighted in red in Fig. \ref{F_tot_OP} and we set $F = 0$. Additionally, note that there are two distinct ($d_{0},q$)-areas of missing data (colored black): one for small ($d_{0},q$)-values, where the initial overlap of the two-soliton system is significant, and thus these cases are excluded, and another for large ($d_{0},q$)-values, where the interaction between solitons is negligible within the fixed simulation time.

 From the presented results, it is clear that in the case of IP configurations the model yields significantly better results as compared to OP configurations. This difference can be attributed to the fact that the soliton overlap remains still small during the dynamics in the repulsive IP configuration, whereas the attractive OP interaction inevitably leads to large overlaps around collision points, in turn invalidating the
 assumptions associated with our model.
 
 For the OP configuration, we observe that the model does a good job at capturing numerous of its qualitative characteristics.
 However, as Fig.~\ref{densities_vs_model_OP} and Fig.~\ref{Nb_vs_model_OP} illustrate, there are scenarios
 where the evolution of the bright atom number exchange is reasonably accurate. In these cases, the particle model
 closely follows the PDE.
In other cases, e.g., Fig.~\ref{Nb_vs_model_OP} (c), the sensitivity
 of the exchange does not enable the accurate prediction thereof and ultimately leads into a failed diagnosis of the
 resulting dynamics. The NS case is again immune to such features,
 due to its conservation of $N_b$and is thus generally well captured. 

\section{Conclusions \& Future Directions} \label{sec:conclusions}
In this study, we investigated the interactions between pairs of DBD and BDB spinor solitons in free space. We explored the dependence of these interactions on the phase differences among the constituent bright solitons and examined the influence of spin interactions on the dynamics. To accomplish this, we created stationary single-soliton solutions which we used as building blocks to initialize the two-soliton system. Based on the symmetries of the system, we identified the different case scenarios and considered the time evolution of the initial two-soliton states for varying initial parameters.

Our numerical results confirm that in-phase (IP) configurations are repulsive, while out-of-phase (OP) configurations are attractive at larger distances but become repulsive at shorter distances, consistent with previous studies
\cite{yanMultipleDarkbrightSolitons2011, katsimigaDarkbrightSolitonInteractions2017}. In the out-of-phase scenario, spin interactions were found to break the bound state, enabling solitons to escape and causing particle exchange among the $m_{F}=0,\pm 1$ components. We furthermore attempted to capture the spinor dynamics by considering an effective classical model based on a Lagrangian approach. The accuracy of the proposed model was tested by direct comparison between the soliton trajectories and the model predictions, employing a fidelity measure based on the comparison between the data and predicted velocities using the sMAPE metric.

Despite its limitations, our proposed effective description successfully captured the dynamical behavior of interacting solitons in the majority of cases, providing a reasonable approximation of the soliton trajectories and the interspecies particle‑exchange dynamics, offering insights into the complex nature of soliton interactions in spinor BECs. The most accurate case was the non-spinorial one,
where atom numbers were conserved. Subsequently, we also generally captured the repulsive interaction
between IP bright solitary waves. The most challenging scenario, due to the exchange of particles
between the components was realized in the setting of OP solitonic interactions.

Future work may focus on refining the model to address its limitations and extend its applicability to more diverse and dynamically sensitive scenarios. In this context, alternative two-soliton Ansätze, based entirely on numerical methods, might better capture the dynamics. Indeed, this problem may constitute
an interesting setting for the potential implementation of discovery techniques of effective ordinary
differential equations via Machine Learning methods, for which numerous techniques exist 
(such as SINDy~\cite{brunton2016discovering}, DeepXDE~\cite{lu2021deepxde}, DUE~\cite{due}, among
many others). Naturally, generalizing also the relevant considerations to higher-dimensional
spinor systems involving vortices and bright solitary waves will be of particular interest in 
its own right.

\section{Acknowledgments}
T. P. acknowledges I. A. Englezos and G. Bougas for useful discussions.
A.R.R. was supported by MCIN/AEI/10.13039/501100011033 and FSE+* from Grants No. JDC2024-055035-I* through the "Juan de la Cierva Fellowship"; No. PID2023-147112NB-C22; No. CNS2022-135529 through the “European Union NextGenerationEU/PRTR”; No.CEX2024-001451-M through the “Unit of Excellence María de Maeztu 2025-2031” Award to the Institute of Cosmos Sciences; and by the Generalitat de Catalunya, Grant No. 021SGR01095.
P.G.K was supported in part by the U.S. National Science Foundation under the award PHY-2408988. This research was partly conducted while P.G.K. was  visiting the Okinawa Institute of Science and
Technology (OIST) through the Theoretical Sciences Visiting Program (TSVP), the University of
Sydney through the visitor program of the Sydney Mathematical Research Institute (SMRI) and the Department of Mechanical Engineering at Seoul National
University through a Fulbright Fellowship. Their support is gratefully acknowledged.
Finally, this work was also  supported by a grant from the Simons Foundation [SFI-MPS-SFM-00011048, P.G.K].

\bibliography{bibliography.bib}

@article{abdullaevDYNAMICSBRIGHTMATTER2012,
  title = {{{DYNAMICS OF BRIGHT MATTER WAVE SOLITONS IN A BOSE}}--{{EINSTEIN CONDENSATE}}},
  author = {Abdullaev, Fatkhulla Kh and Gammal, Arnaldo and Kamchatnov, Anatoly M. and Tomio, Lauro},
  year = {2012},
  month = jan,
  journal = {Int. J. Mod. Phys. B},
  publisher = {World Scientific Publishing Company},
  doi = {10.1142/S0217979205032279},
  urldate = {2024-06-19},
  langid = {english}
}

@article{due,
author = {Chen, Junfeng and Wu, Kailiang and Xiu, Dongbin},
title = {DUE: A Deep Learning Framework and Library for Modeling Unknown Equations},
journal = {SIAM Review},
volume = {67},
number = {4},
pages = {873-902},
year = {2025},
doi = {10.1137/24M1671827},

URL = { 
    
        
    
    

},
eprint = { 
    
        
    
    

}
}

@article{brunton2016discovering,
  title={Discovering governing equations from data by sparse identification of nonlinear dynamical systems},
  author={Brunton, Steven L and Proctor, Joshua L and Kutz, J Nathan},
  journal={Proc. Nat. Acad. of Sci.},
  volume={113},
  number={15},
  pages={3932--3937},
  year={2016},
  publisher={National Acad Sciences}
}

@article{lu2021deepxde,
  title={Deep{XDE}: A deep learning library for solving differential equations},
  author={Lu, L. and Meng, X. and Mao, Z. and Karniadakis, G.E.},
  journal={SIAM Review},
  volume={63},
  number={1},
  pages={208--228},
  year={2021},
  publisher={SIAM}
}

@article{sheppard,
  title = {Polarized dark solitons in isotropic Kerr media},
  author = {Sheppard, Adrian P. and Kivshar, Yuri S.},
  journal = {Phys. Rev. E},
  volume = {55},
  issue = {4},
  pages = {4773--4782},
  numpages = {0},
  year = {1997},
  month = {Apr},
  publisher = {American Physical Society},
  doi = {10.1103/PhysRevE.55.4773},
  url = {https://link.aps.org/doi/10.1103/PhysRevE.55.4773}
}

@article{mossman2024observation,
  author  = {Mossman, Sean M. and Katsimiga, Garyfallia C. and Mistakidis, Simeon I. and Romero-Ros, Alejandro and Bersano, Thomas M. and Schmelcher, Peter and Kevrekidis, Panayotis G. and Engels, Peter},
  title   = {Observation of dense collisional soliton complexes in a two-component Bose-Einstein condensate},
  journal = {Commun. Phys.},
  volume  = {7},
  pages   = {163},
  year    = {2024},
  doi     = {10.1038/s42005-024-01659-w}
}

@misc{tamura2025observationmanybodycoherencequasionedimensional,
      title={}, 
      author={Hikaru Tamura and Sambit Banerjee and Rongjie Li and Panayotis Kevrekidis and Simeon I. Mistakidis and Chen-Lung Hung},
      year={2025},
      eprint={2506.13597},
      archivePrefix={arXiv},
      primaryClass={},
      url={https://arxiv.org/abs/2506.13597}, 
}

@article{beckerOscillationsInteractionsDark2008,
  title = {Oscillations and Interactions of Dark and Dark--Bright Solitons in {{Bose}}--{{Einstein}} Condensates},
  author = {Becker, Christoph and Stellmer, Simon and {Soltan-Panahi}, Parvis and D{\"o}rscher, S{\"o}ren and Baumert, Mathis and Richter, Eva-Maria and Kronj{\"a}ger, Jochen and Bongs, Kai and Sengstock, Klaus},
  year = {2008},
  month = jun,
  journal = {Nature Phys.},
  volume = {4},
  number = {6},
  pages = {496--501},
  publisher = {Nature Publishing Group},
  issn = {1745-2481},
  doi = {10.1038/nphys962},
  urldate = {2024-06-19},
  copyright = {2008 Springer Nature Limited},
  langid = {english}
}

@article{bersanoThreeComponentSolitonStates2018,
  title = {Three-{{Component Soliton States}} in {{Spinor}} \${{F}}=1\$ {{Bose-Einstein Condensates}}},
  author = {Bersano, T. M. and Gokhroo, V. and Khamehchi, M. A. and D'Ambroise, J. and Frantzeskakis, D. J. and Engels, P. and Kevrekidis, P. G.},
  year = {2018},
  month = feb,
  journal = {Phys. Rev. Lett.},
  volume = {120},
  number = {6},
  pages = {063202},
  publisher = {American Physical Society},
  doi = {10.1103/PhysRevLett.120.063202},
  urldate = {2024-06-20}
}

@article{changCoherentSpinorDynamics2005,
  title = {Coherent Spinor Dynamics in a Spin-1 {{Bose}} Condensate},
  author = {Chang, Ming-Shien and Qin, Qishu and Zhang, Wenxian and You, Li and Chapman, Michael S.},
  year = {2005},
  month = nov,
  journal = {Nature Phys},
  volume = {1},
  number = {2},
  pages = {111--116},
  publisher = {Nature Publishing Group},
  issn = {1745-2481},
  doi = {10.1038/nphys153},
  urldate = {2024-06-19},
  copyright = {2005 Springer Nature Limited},
  langid = {english}
}

@article{changObservationSpinorDynamics2004,
  title = {Observation of {{Spinor Dynamics}} in {{Optically Trapped}} \${\textasciicircum}\{87\}{\textbackslash}mathrm\{\vphantom\}{{Rb}}\vphantom\{\}\$ {{Bose-Einstein Condensates}}},
  author = {Chang, M.-S. and Hamley, C. D. and Barrett, M. D. and Sauer, J. A. and Fortier, K. M. and Zhang, W. and You, L. and Chapman, M. S.},
  year = {2004},
  month = apr,
  journal = {Phys. Rev. Lett.},
  volume = {92},
  number = {14},
  pages = {140403},
  publisher = {American Physical Society},
  doi = {10.1103/PhysRevLett.92.140403},
  urldate = {2024-06-19}
}

@article{christodoulidesBlackWhiteVector1988,
  title = {Black and White Vector Solitons in Weakly Birefringent Optical Fibers},
  author = {Christodoulides, D. N.},
  year = {1988},
  month = oct,
  journal = {Phys. Lett. A},
  volume = {132},
  number = {8},
  pages = {451--452},
  issn = {0375-9601},
  doi = {10.1016/0375-9601(88)90511-7},
  urldate = {2024-06-19}
}

@article{davisBoseEinsteinCondensationGas1995,
  title = {Bose-{{Einstein Condensation}} in a {{Gas}} of {{Sodium Atoms}}},
  author = {Davis, K. B. and Mewes, M. -O. and Andrews, M. R. and {van Druten}, N. J. and Durfee, D. S. and Kurn, D. M. and Ketterle, W.},
  year = {1995},
  month = nov,
  journal = {Phys. Rev. Lett.},
  volume = {75},
  number = {22},
  pages = {3969--3973},
  publisher = {American Physical Society},
  doi = {10.1103/PhysRevLett.75.3969},
  urldate = {2024-06-19}
}

@article{farolfiObservationMagneticSolitons2020,
  title = {Observation of {{Magnetic Solitons}} in {{Two-Component Bose-Einstein Condensates}}},
  author = {Farolfi, A. and Trypogeorgos, D. and Mordini, C. and Lamporesi, G. and Ferrari, G.},
  year = {2020},
  month = jul,
  journal = {Phys. Rev. Lett.},
  volume = {125},
  number = {3},
  pages = {030401},
  publisher = {American Physical Society},
  doi = {10.1103/PhysRevLett.125.030401},
  urldate = {2024-06-19}
}

@article{frantzeskakisDarkSolitonsAtomic2010,
  title = {Dark Solitons in Atomic {{Bose}}--{{Einstein}} Condensates: From Theory to Experiments},
  shorttitle = {Dark Solitons in Atomic {{Bose}}--{{Einstein}} Condensates},
  author = {Frantzeskakis, D. J.},
  year = {2010},
  month = feb,
  journal = {J. Phys. A: Math. Theor.},
  volume = {43},
  number = {21},
  pages = {213001},
  issn = {1751-8121},
  doi = {10.1088/1751-8113/43/21/213001},
  urldate = {2024-06-19},
  langid = {english}
}

@article{hamnerGenerationDarkBrightSoliton2011,
  title = {Generation of {{Dark-Bright Soliton Trains}} in {{Superfluid-Superfluid Counterflow}}},
  author = {Hamner, C. and Chang, J. J. and Engels, P. and Hoefer, M. A.},
  year = {2011},
  month = feb,
  journal = {Phys. Rev. Lett.},
  volume = {106},
  number = {6},
  pages = {065302},
  publisher = {American Physical Society},
  doi = {10.1103/PhysRevLett.106.065302},
  urldate = {2024-06-19}
}

@article{iedaExactSolitonSolutions2006,
  title = {Exact Soliton Solutions of Spinor {{Bose-Einstein}} Condensates},
  author = {Ieda, J. and Miyakawa, T. and Wadati, M.},
  year = {2006},
  month = apr,
  journal = {Las. Phys.},
  volume = {16},
  number = {4},
  pages = {678--682},
  issn = {1555-6611},
  doi = {10.1134/S1054660X06040220},
  urldate = {2024-06-19},
  langid = {english}
}

@article{iedaMatterWaveSolitonsSpinor2004,
  title = {Matter-{{Wave Solitons}} in an {{F}}=1 {{Spinor Bose}}--{{Einstein Condensate}}},
  author = {Ieda, Jun'ichi and Miyakawa, Takahiko and Wadati, Miki},
  year = {2004},
  month = nov,
  journal = {J. Phys. Soc. Jpn.},
  volume = {73},
  number = {11},
  pages = {2996--3007},
  publisher = {The Physical Society of Japan},
  issn = {0031-9015},
  doi = {10.1143/JPSJ.73.2996},
  urldate = {2024-06-19}
}

@article{katsimigaDarkbrightSolitonInteractions2017,
  title = {Dark-Bright Soliton Interactions beyond the Integrable Limit},
  author = {Katsimiga, G. C. and Stockhofe, J. and Kevrekidis, P. G. and Schmelcher, P.},
  year = {2017},
  month = jan,
  journal = {Phys. Rev. A},
  volume = {95},
  number = {1},
  pages = {013621},
  publisher = {American Physical Society},
  doi = {10.1103/PhysRevA.95.013621},
  urldate = {2024-06-09}
}

@article{katsimigaDarkbrightSolitonPairs2018,
  title = {Dark-Bright Soliton Pairs: {{Bifurcations}} and Collisions},
  shorttitle = {Dark-Bright Soliton Pairs},
  author = {Katsimiga, G. C. and Kevrekidis, P. G. and Prinari, B. and Biondini, G. and Schmelcher, P.},
  year = {2018},
  month = apr,
  journal = {Phys. Rev. A},
  volume = {97},
  number = {4},
  pages = {043623},
  publisher = {American Physical Society},
  doi = {10.1103/PhysRevA.97.043623},
  urldate = {2024-06-19}
}

@article{katsimigaPhaseDiagramStability2021,
  title = {Phase Diagram, Stability and Magnetic Properties of Nonlinear Excitations in Spinor {{Bose}}--{{Einstein}} Condensates},
  author = {Katsimiga, G. C. and Mistakidis, S. I. and Schmelcher, P. and Kevrekidis, P. G.},
  year = {2021},
  month = oct,
  journal = {New J. Phys.},
  volume = {23},
  number = {1},
  pages = {013015},
  publisher = {IOP Publishing},
  issn = {1367-2630},
  doi = {10.1088/1367-2630/abd27c},
  urldate = {2024-06-09},
  langid = {english}
}

@article{katsimigaStabilityDynamicsDarkBright2017,
  title = {Stability and {{Dynamics}} of {{Dark-Bright Soliton Bound States Away}} from the {{Integrable Limit}}},
  author = {Katsimiga, Garyfallia C. and Stockhofe, Jan and Kevrekidis, Panagiotis G. and Schmelcher, Peter},
  year = {2017},
  month = apr,
  journal = {Appl. Sci.},
  volume = {7},
  number = {4},
  pages = {388},
  publisher = {Multidisciplinary Digital Publishing Institute},
  issn = {2076-3417},
  doi = {10.3390/app7040388},
  urldate = {2024-06-09},
  copyright = {http://creativecommons.org/licenses/by/3.0/},
  langid = {english}
}

@article{kawaguchiSpinorBoseEinstein2012,
  title = {Spinor {{Bose}}--{{Einstein}} Condensates},
  author = {Kawaguchi, Yuki and Ueda, Masahito},
  year = {2012},
  month = nov,
  journal = {Phys. Rep.},
  series = {},
  volume = {520},
  number = {5},
  pages = {253--381},
  issn = {0370-1573},
  doi = {10.1016/j.physrep.2012.07.005},
  urldate = {2024-06-19}
}

@book{kelleySolvingNonlinearEquations2003,
  title = {Solving {{Nonlinear Equations}} with {{Newton}}'s {{Method}}},
  author = {Kelley, C. T.},
  year = {2003},
  month = jan,
  series = {Fundamentals of {{Algorithms}}},
  publisher = {{Society for Industrial and Applied Mathematics}},
  doi = {10.1137/1.9780898718898},
  urldate = {2024-06-19},
  isbn = {978-0-89871-546-0}
}

@article{kevrekidisSolitonsCoupledNonlinear2016,
  title = {Solitons in Coupled Nonlinear {{Schr{\"o}dinger}} Models: {{A}} Survey of Recent Developments},
  shorttitle = {Solitons in Coupled Nonlinear {{Schr{\"o}dinger}} Models},
  author = {Kevrekidis, P. G. and Frantzeskakis, D. J.},
  year = {2016},
  month = nov,
  journal = {Reviews in Physics},
  volume = {1},
  pages = {140--153},
  issn = {2405-4283},
  doi = {10.1016/j.revip.2016.07.002},
  urldate = {2024-06-19}
}

@article{kivsharLagrangianApproachDark1995,
  title = {Lagrangian Approach for Dark Solitons},
  author = {Kivshar, Yuri S. and Kr{\'o}likowski, Wies{\l}aw},
  year = {1995},
  month = feb,
  journal = {Optics Commun.},
  volume = {114},
  number = {3},
  pages = {353--362},
  issn = {0030-4018},
  doi = {10.1016/0030-4018(94)00644-A},
  urldate = {2025-06-19}
}

@article{lannigCollisionsThreeComponentVector2020,
  title = {Collisions of {{Three-Component Vector Solitons}} in {{Bose-Einstein Condensates}}},
  author = {Lannig, Stefan and Schmied, Christian-Marcel and Pr{\"u}fer, Maximilian and Kunkel, Philipp and Strohmaier, Robin and Strobel, Helmut and Gasenzer, Thomas and Kevrekidis, Panayotis G. and Oberthaler, Markus K.},
  year = {2020},
  month = oct,
  journal = {Phys. Rev. Lett.},
  volume = {125},
  number = {17},
  pages = {170401},
  publisher = {American Physical Society},
  doi = {10.1103/PhysRevLett.125.170401},
  urldate = {2024-06-19}
}

@article{middelkampDynamicsDarkBright2011,
  title = {Dynamics of Dark--Bright Solitons in Cigar-Shaped {{Bose}}--{{Einstein}} Condensates},
  author = {Middelkamp, S. and Chang, J. J. and Hamner, C. and {Carretero-Gonz{\'a}lez}, R. and Kevrekidis, P. G. and Achilleos, V. and Frantzeskakis, D. J. and Schmelcher, P. and Engels, P.},
  year = {2011},
  month = jan,
  journal = {Physics Letters A},
  volume = {375},
  number = {3},
  pages = {642--646},
  issn = {0375-9601},
  doi = {10.1016/j.physleta.2010.11.025},
  urldate = {2024-06-19}
}

@article{nguyenCollisionsMatterwaveSolitons2014,
  title = {Collisions of Matter-Wave Solitons},
  author = {Nguyen, Jason H. V. and Dyke, Paul and Luo, De and Malomed, Boris A. and Hulet, Randall G.},
  year = {2014},
  month = dec,
  journal = {Nature Phys.},
  volume = {10},
  number = {12},
  pages = {918--922},
  publisher = {Nature Publishing Group},
  issn = {1745-2481},
  doi = {10.1038/nphys3135},
  urldate = {2024-06-19},
  copyright = {2014 Springer Nature Limited},
  langid = {english}
}

@article{nistazakisBrightdarkSolitonComplexes2008,
  title = {Bright-Dark Soliton Complexes in Spinor {{Bose-Einstein}} Condensates},
  author = {Nistazakis, H. E. and Frantzeskakis, D. J. and Kevrekidis, P. G. and Malomed, B. A. and {Carretero-Gonz{\'a}lez}, R.},
  year = {2008},
  month = mar,
  journal = {Phys. Rev. A},
  volume = {77},
  number = {3},
  pages = {033612},
  publisher = {American Physical Society},
  doi = {10.1103/PhysRevA.77.033612},
  urldate = {2024-06-19}
}

@article{nistazakisPolarizedStatesDomain2007,
  title = {Polarized States and Domain Walls in Spinor {{Bose-Einstein}} Condensates},
  author = {Nistazakis, H. E. and Frantzeskakis, D. J. and Kevrekidis, P. G. and Malomed, B. A. and {Carretero-Gonz{\'a}lez}, R. and Bishop, A. R.},
  year = {2007},
  month = dec,
  journal = {Phys. Rev. A},
  volume = {76},
  number = {6},
  pages = {063603},
  publisher = {American Physical Society},
  doi = {10.1103/PhysRevA.76.063603},
  urldate = {2024-06-19}
}

@article{ostrovskayaNonlinearTheorySolitoninduced1998,
  title = {Nonlinear Theory of Soliton-Induced Waveguides},
  author = {Ostrovskaya, Elena A. and Kivshar, Yuri S.},
  year = {1998},
  month = aug,
  journal = {Opt. Lett.},
  volume = {23},
  number = {16},
  pages = {1268--1270},
  publisher = {Optica Publishing Group},
  issn = {1539-4794},
  doi = {10.1364/OL.23.001268},
  urldate = {2024-06-19},
  copyright = {{\copyright} 1998 Optical Society of America},
  langid = {english}
}

@book{pitaevskiiBoseEinsteinCondensationSuperfluidity2016,
  title = {Bose-{{Einstein Condensation}} and {{Superfluidity}}},
  author = {Pitaevskii, Lev and Stringari, Sandro},
  year = {2016},
  month = jan,
  publisher = {Oxford University Press},
  doi = {10.1093/acprof:oso/9780198758884.001.0001},
  urldate = {2024-06-19},
  isbn = {978-0-19-181872-1},
  langid = {english}
}

@article{romero-rosControlledGenerationDarkbright2019,
  title = {Controlled Generation of Dark-Bright Soliton Complexes in Two-Component and Spinor {{Bose-Einstein}} Condensates},
  author = {{Romero-Ros}, A. and Katsimiga, G. C. and Kevrekidis, P. G. and Schmelcher, P.},
  year = {2019},
  month = jul,
  journal = {Phys. Rev. A},
  volume = {100},
  number = {1},
  pages = {013626},
  publisher = {American Physical Society},
  doi = {10.1103/PhysRevA.100.013626},
  urldate = {2024-06-19}
}

@article{stamper-kurnSpinorBoseGases2013,
  title = {Spinor {{Bose}} Gases: {{Symmetries}}, Magnetism, and Quantum Dynamics},
  shorttitle = {Spinor {{Bose}} Gases},
  author = {{Stamper-Kurn}, Dan M. and Ueda, Masahito},
  year = {2013},
  month = jul,
  journal = {Rev. Mod. Phys.},
  volume = {85},
  number = {3},
  pages = {1191--1244},
  publisher = {American Physical Society},
  doi = {10.1103/RevModPhys.85.1191},
  urldate = {2024-06-19}
}

@article{stellmerCollisionsDarkSolitons2008,
  title = {Collisions of {{Dark Solitons}} in {{Elongated Bose-Einstein Condensates}}},
  author = {Stellmer, S. and Becker, C. and {Soltan-Panahi}, P. and Richter, E.-M. and D{\"o}rscher, S. and Baumert, M. and Kronj{\"a}ger, J. and Bongs, K. and Sengstock, K.},
  year = {2008},
  month = sep,
  journal = {Phys. Rev. Lett.},
  volume = {101},
  number = {12},
  pages = {120406},
  publisher = {American Physical Society},
  doi = {10.1103/PhysRevLett.101.120406},
  urldate = {2024-06-19}
}

@article{trilloOpticalSolitaryWaves1988,
  title = {Optical Solitary Waves Induced by Cross-Phase Modulation},
  author = {Trillo, S. and Wabnitz, S. and Wright, E. M. and Stegeman, G. I.},
  year = {1988},
  month = oct,
  journal = {Opt. Lett.},
  volume = {13},
  pages = {871--873},
  publisher = {Optica Publishing Group},
  issn = {1539-4794},
  doi = {10.1364/OL.13.000871},
  urldate = {2024-06-19},
  copyright = {{\copyright} 1988 Optical Society of America},
  langid = {english}
}

@article{tsuchidaCoupledModifiedKortewegde1998,
  title = {The {{Coupled Modified Korteweg-de Vries Equations}}},
  author = {Tsuchida, Takayuki and Wadati, Miki},
  year = {1998},
  month = apr,
  journal = {J. Phys. Soc. Jpn.},
  volume = {67},
  number = {4},
  pages = {1175--1187},
  publisher = {The Physical Society of Japan},
  issn = {0031-9015},
  doi = {10.1143/JPSJ.67.1175},
  urldate = {2024-06-19}
}

@article{wellerExperimentalObservationOscillating2008,
  title = {Experimental {{Observation}} of {{Oscillating}} and {{Interacting Matter Wave Dark Solitons}}},
  author = {Weller, A. and Ronzheimer, J. P. and Gross, C. and Esteve, J. and Oberthaler, M. K. and Frantzeskakis, D. J. and Theocharis, G. and Kevrekidis, P. G.},
  year = {2008},
  month = sep,
  journal = {Phys. Rev. Lett.},
  volume = {101},
  number = {13},
  pages = {130401},
  publisher = {American Physical Society},
  doi = {10.1103/PhysRevLett.101.130401},
  urldate = {2024-06-19}
}

@article{xiongDynamicalCreationComplex2010,
  title = {Dynamical Creation of Complex Vector Solitons in Spinor {{Bose-Einstein}} Condensates},
  author = {Xiong, Bo and Gong, Jiangbin},
  year = {2010},
  month = mar,
  journal = {Phys. Rev. A},
  volume = {81},
  number = {3},
  pages = {033618},
  publisher = {American Physical Society},
  doi = {10.1103/PhysRevA.81.033618},
  urldate = {2024-06-19}
}

@article{yanMultipleDarkbrightSolitons2011,
  title = {Multiple Dark-Bright Solitons in Atomic {{Bose-Einstein}} Condensates},
  author = {Yan, D. and Chang, J. J. and Hamner, C. and Kevrekidis, P. G. and Engels, P. and Achilleos, V. and Frantzeskakis, D. J. and {Carretero-Gonz{\'a}lez}, R. and Schmelcher, P.},
  year = {2011},
  month = nov,
  journal = {Phys. Rev. A},
  volume = {84},
  number = {5},
  pages = {053630},
  publisher = {American Physical Society},
  doi = {10.1103/PhysRevA.84.053630},
  urldate = {2024-06-19}
}

@article{shuklaSaltRockCreep2025,
  title = {Salt Rock Creep Deformation Forecasting Using Deep Neural Networks and Analytical Models for Subsurface Energy Storage Applications},
  author = {Shukla, Pradeep Kumar and Chakraborty, Tanujit and Sari, Mustafa and Sarout, Joel and Mandal, Partha Pratim},
  year = {2025},
  month = oct,
  journal = {Sci. Rep.},
  volume = {15},
  number = {1},
  pages = {34560},
  publisher = {Nature Publishing Group},
  issn = {2045-2322},
  doi = {10.1038/s41598-025-17940-z},
  urldate = {2026-06-01},
  copyright = {2025 The Author(s)},
  langid = {english}
}

@article{PhysRevA.75.023617,
  title = {Multicomponent gap solitons in spinor Bose-Einstein condensates},
  author = {Dabrowska-Wuster, Beata J. and Ostrovskaya, Elena A. and Alexander, Tristram J. and Kivshar, Yuri S.},
  journal = {Phys. Rev. A},
  volume = {75},
  number = {2},
  pages = {023617},
  year = {2007},
  doi = {10.1103/PhysRevA.75.023617}
}
\bibliographystyle{apsrev4-1}
\end{document}